\documentclass[12pt]{article}

\newcommand{\be}{\begin{equation}}
\newcommand{\ee}{\end{equation}}
\newcommand{\bea}{\begin{eqnarray}}
\newcommand{\eea}{\end{eqnarray}}

\newcommand{\bit}{\begin{itemize}}
\newcommand{\eit}{\end{itemize}}
\newcommand{\no}{\noindent}

\begin{document}
{\sf \title{The geometry of wrapped M5-branes in Calabi-Yau 2-folds}
\author{Ansar Fayyazuddin\footnote{email: Ansar\_ Fayyazuddin@baruch.cuny.edu} $^1$, Tasneem Zehra Husain\footnote{email: tasneem@physics.harvard.edu} $^2$ and Ioanna Pappa\footnote{gianna@physto.se}$^{\;\,3,4}$}
\maketitle
\begin{center}
\vspace{-1cm}
{\it $^1$ Department of Natural Sciences, Baruch College, \\
City University of New York, New York, NY\\ 
$^2$ Jefferson Physical Laboratory, Harvard University, \\
Cambridge, MA 02138\\
$^3$ Department of Physics, Stockholm University, \\
Stockholm SE 106 91 , Sweden\\
$^4$Physics Department, National Technical University of Athens\\
Zografou Campus, GR 157 80, Athens, Greece.}
\end{center}

\begin{abstract}
We study the geometry of M5-branes wrapping a 2-cycle  which is Special Lagrangian with respect to a specific complex structure in a Calabi-Yau two-fold.  Using methods recently applied to the three-fold case, we are again able find a characterization of the geometry, in terms of a non-integrable almost complex structure and a (2,0) form. This time, however, due to the hyper-K{\"a}hler nature of the underlying 2-fold we also have the freedom of choosing a different almost complex structure with respect to which the wrapped 2-cycle is holomorphic.  We show that this latter almost complex structure is integrable.  We then relate our geometry to previously found geometries of M5-branes wrapping holomophic cycles and go further to prove some previously unknown results for M5-branes on holomorphic cycles.  
\end{abstract}

\vspace{-19cm}
\begin{flushright}
HUTP-05/A0040 \\
BCCUNY-HEP /05-04 \\
hep-th/xxxxxxx
\end{flushright}

\thispagestyle{empty}

\newpage

\tableofcontents

\section{Introduction}
In a recent paper \cite{fh}, two of us studied M5-branes wrapping Special Lagrangian 3-cycles in Calabi-Yau three-folds.  We argued that the almost complex structure of the underlying Calabi-Yau three-fold survives the wrapping.  With this assumption we were able to characterize the geometry of the wrapped branes in terms of the almost complex structure as well as a (3,0) form.  

In the present paper we apply our methods to the case of M5-branes wrapping Special Lagrangian 2-cycles in Calabi-Yau 2-folds.  By Calabi-Yau 2-folds we will mean compact or non-compact Ricci flat complex manifolds of dimension 2.  Thus $K3, T^4, C^4$ and  $ALE$-spaces are all examples of two-folds for which our methods are valid.  

What makes the case of two-folds particularly interesting (and simple) is that Calabi-Yau two-folds are hyper-K{\" a}hler. This means that there are an $SU(2)$ worth of incompatible complex structures with respect to which the two-fold is a K{\"a}hler manifold.  In fact the choice of complex structure is broad enough so that any supersymmetric 2-cycle is holomorphic with respect to one of the allowed complex structures.  We will exploit this property later as a test of our methods.

The first part of the present paper is a study of the geometry of M5-branes wrapping Special Lagrangian cycles in two-folds.  We find as in \cite{fh} that the supergravity background is expressed in terms of a (2,0) form $\Omega$ as well as a distinguished (1,1) form which we call $B$.  We then show how to re-express our results in a different and, as it turns out, integrable almost complex structure.  Here we make contact with previous work on M5-branes wrapping holomorphic cycles \cite{danda}.  In section 5 we discuss calibrations in the geometry produced by wrapped M5-branes and give some of our constraints a physical interpretation.  We end with conclusions in section 6.

\section{Special Lagrangian Cycles and their Killing Spinors}

In this section we characterize the Killing spinor for branes wrapping Special Lagrangian 2-cycles in Calabi-Yau two-folds.  Our discussion here follows closely \cite{fh}.
Calabi-Yau two-folds come equipped with a Kahler form $\omega$ which is a 
distinguished member of $H^{(1,1)}$ as well as a unique holomorphic 
(2,0)-form $\Omega$.  A Special Lagrangian sub-manifold, $\Sigma$, 
of a Calabi-Yau two-fold is defined by the following set of conditions:
\begin{eqnarray}
\omega |_{\Sigma} &=& 0 \nonumber \\
\Re (e^{i\theta_\Sigma} \Omega) |_{\Sigma} &=& d\mbox{vol}(\Sigma)
\end{eqnarray}
where $\theta_{\Sigma}$ is a $\Sigma$-dependent constant phase.  
In other words, the pullback of the Kahler form vanishes on $\Sigma$ and, 
up to a phase, the pullback of $\Omega$ is the volume form on $\Sigma$.  
In fact, $\Omega$ 
gives a BPS bound on 2-cycles so that only ones which saturate this bound 
are minimum volume cycles in their homology class. $\Omega$, thus, 
is a calibrating form.  

Given a SpelL $\Sigma$ we would like to find the 
fraction of supersymmetry preserved by an M5-brane wrapping it.  
The Killing spinors of a $p$-brane with embedding coordinates $X^{A}$ 
satisfy the following projection condition \cite{BBS}:
\begin{equation}
\epsilon = \frac{1}{p!} \epsilon^{\alpha_0\alpha_1 ... \alpha_p}
\Gamma_{A_0A_1 ... A_p} \partial_{\alpha_0} X^{A_0}
\partial_{\alpha_1} X^{A_1} ....
\partial_{\alpha_p} X^{A_p} \epsilon
\label{KSeqn}
\end{equation}
where $\epsilon$ is a Majorana spinor in 11 dimensions.  

Choosing static gauge along the directions $0123$, the condition on the 
Killing spinors $\epsilon$ for an M5-brane wrapping a Special Lagrangian 
2-cycle, is given by (\ref{KSeqn}): 
\begin{equation}
\epsilon^{ab}
\Gamma_{0 1 2 3} \Gamma_{i j }
\partial_{a} X^{i} \partial_{b} X^{j}\epsilon =
\epsilon
\label{dpsi}
\end{equation}
where ${\sigma}^a, {\sigma}^b$ are coordinates on the Special 
Lagrangian two-cycle and the $X^i$ are coordinates in the Calabi-Yau. 
Using complex coordinates on the Calabi-Yau, the projection condition becomes:
\be
\frac{1}{2}\epsilon^{abc}
\Gamma_{0 1 2 3}[ \Gamma_{MN}
\partial_{a} Z^M \partial_{b} Z^N  +  2\Gamma_{{M}\bar{N}}
\partial_{a} Z^{M} \partial_{b} Z^{\bar N} + \Gamma_{\bar{M}\bar{N}}
\partial_{a} Z^{\bar M} \partial_{b} Z^{\bar N}]\epsilon =
\epsilon
\ee 
Since a Calabi-Yau is a complex manifold there is a choice of frame 
reflecting the complex structure - so that the only non-zero frame vectors are those of  
the form $e^a_A$ where $a, A$ are either both holomorphic or both 
anti-holomorphic indices.  The frame is defined as usual so that the metric 
on the Calabi-Yau can be expressed as $G_{M\bar{N}} = \eta_{m\bar n}e^m_Me^{\bar n}_{\bar{N}}$.  

If we define $\gamma_m = e_m^M\Gamma_M$ for $m=1,2$, these gamma matrices 
satisfy the flat space Clifford algebra which is identical, as is well known, 
to the creation annihilation algebra for 2 families of fermions.  
Thus a spinor can be represented by a set of states in a Fock space.  
We define the vacuum state to be $ \epsilon_{00}$ and declare 
$\gamma_m$ to be annihilation operators, so that $\gamma_m \epsilon_{00} =0$. The remaining states can then be labeled by their occupation numbers corresponding to the action of the gamma matrices. We will use this construction to express the Killing spinor as a linear combination of Fock space states.

We introduce $U, V$ as holomorphic coordinates on the Calabi-Yau.  Since the pullback of the (2,0)-form $\Omega$ is the 
volume form on the SpelL: 
$$\epsilon^{ab}\epsilon_{mn}e^m_Ue^n_V\partial_{a} U \partial_{b} V = 1$$ we can impose the condition (\ref{dpsi}) to find that generically\footnote{As in the three-fold case \cite{fh} the only spinors that survive compactification are the holonomy singlet spinors, which in this case are the $SU(2)$ singlets $\epsilon_{00}$ and $\epsilon_{11}$.  When considering $T^4$ or $C^2$ the same spinors survive as long as the SpelL 2-cycle is non-trivially embedded in the space.  M5-branes wrapping trivially embedded SpelLs are flat M5-branes, which are well studied and will form a sub-case of the more general story pursued here.}:
\begin{equation}
\epsilon_{01} = \epsilon_{10} =  0
\end{equation}
\no
The only components that survive are
$\epsilon_{00}$ and $\epsilon_{11}$, and these must obey
\bea
\gamma_{0123}\gamma_{u v } \epsilon_{11} 
&=& \epsilon_{00} \nonumber \\
\gamma_{0123}\gamma_{{\bar u} {\bar v}} \epsilon_{00}
&=& \epsilon_{11}
\label{susyconst}
\eea

\no
It is convenient to pick the flat gamma matrices so that $\gamma_a^*=-\gamma_a$ for $a=0,\dots,9$ and $\gamma_{(10)} =\gamma_0\cdots\gamma_9$ is real.  The Majorana condition on $\epsilon$ expressed in this gamma matrix basis is $\gamma_{(10)}\epsilon = \epsilon^*$ or:
\be
\gamma_{(10)} \epsilon_{00}^* = \epsilon_{11}  \;\;\;\;\;\;\;\;\;\;\;\;\;\;
\gamma_{(10)} \epsilon_{00}^* = \epsilon_{11}
\ee

\no
The wrapped M5-brane then preserves $\frac{1}{4}$ of the spacetime
supersymmetry, corresponding to 8 real degrees of freedom or N=2 in four dimensions.  The Killing spinor is then:
\be
\epsilon = \epsilon_{00} + \epsilon_{11} \label{epsilon}
\ee
with $\epsilon_{00}, \epsilon_{11}$ satisfying the above constraints.
By using the identity $\gamma_0\dots\gamma_{(10)}=1$ we can show that
\begin{equation}
\gamma_{0123}\gamma_{89(10)}\epsilon = -\epsilon.
\end{equation}

\section{Supergravity solutions for wrapped branes}
\label{ansatz}
\no
\subsection{Setting up}
We are interested in solving the supergravity equations of motion for M5-branes wrapping a Special Lagrangian 2-cycle which preserves the fraction (\ref{epsilon}) of supersymmetry.  To this end we will use the methods of our previous paper \cite{fh} in which we interpret (\ref{epsilon}) as the supersymmetry variation parameter in the Killing spinor equation of 11-dimensional supergravity.  
We argued in \cite{fh} that although $\epsilon$ was initially defined using the complex structure of the underlying Calabi-Yau whose 2-cycle the M5-brane is wrapping, we only need an {\it almost} complex structure to define $\epsilon$.  Thus, as in \cite{fh}, we will assume that the full geometry produced by the wrapped branes continues to have an almost complex structure in the 4-dimensional part that was initially the Calabi-Yau two-fold.  We now develop the formalism as it applies to the case at hand.

Our starting point is M-theory compactified on a Calabi-Yau two-fold with no other fields turned on.  The initial geometry is then R$^{(6,1)}\times$CY$_2$.  Consider wrapping an M5-brane on a 2-cycle in the two-fold.  The M5-brane then has 4 flat directions oriented along $0123$ and 2 directions wrapped on a Special Lagrangian 2-cycle.  The minimum isometries of the geometry are then $SO(3,1)\times SO(3)\subset SO(6,1)$ which is the unbroken subgroup of the isometries of R$^{(6,1)}$.  The geometry may enjoy additional isometries depending on the CY$_2$ in question.  A metric ansatz with these isometries is then:
\be
ds^2 = H_1^2\eta_{\mu\nu}dx^\mu dx^\nu + g_{IJ}dy^Idy^J + H_2^2 \delta_{\alpha\beta}dx^\alpha dx^\beta.
\ee  
Here $\mu , \nu = 0,1,2,3$ are the directions along the M5-brane transverse to the Calabi-Yau two-fold,  $\alpha, \beta =8,9, 10$ are the directions transverse to the M5-brane and the two-fold.  Finally, $y^I, I,J= 1,\dots, 4$ are directions along what was initially the two-fold. 

The requirement that there is an almost complex structure inherited from the underlying Calabi-Yau two-fold means that there is a basis of 1-forms $e^m$ with $m=u, v$ which are classified by the almost complex structure as $(1,0)$ forms.  Their complex conjugates $e^{\bar{m}}$ are then $(0,1)$ forms.  We will pick these 1-forms in such a way that they satisfy 
\be
g_{IJ} = \eta_{m\bar{n}}(e^m_Ie^{\bar{n}}_J + e^m_Je^{\bar{n}}_I)
\ee 
where $\eta_{m\bar{n}} = 1/2 \delta_{mn}$ is the flat space metric.  We preserve the metric and the notion of $(1,0)$ and $(0,1)$ forms under $U(1)\times SU(2)$ rotations that transform $e^m$ as a doublet under the $SU(2)$ and charge $1$ under the $U(1)$, and $e^{\bar{m}}$ in the complex conjugate representation.  Another singlet under this $U(1)\times SU(2)$ group can be constructed as 
\be
B = e^u\wedge e^{\bar u} + e^v\wedge e^{\bar v}.
\ee  
A third combination:
\be
\Omega = e^u\wedge e^v
\ee
is explicitly $SU(2)$ invariant but is not $U(1)$ invariant.  $\Omega$ is a $(2,0)$ form in this classification which, a priori, need not be globally defined (even when the almost complex structure is), since it is not $U(1)$ invariant .  As we shall see, however, physical fields will be expressed in terms of $\Omega$ requiring that this must in fact be a globally defined object.

In addition to the metric, 11-dimensional supergravity has a 4-form field strength $F$.  To preserve the $SO(3,1)\times SO(3)$ isometries, we can only turn on the following components of $F$: $F_{IJ\alpha\beta}, F_{I89(10)}, F_{0123}$.  We will explicitly set $F_{0123}=0$ in the remainder of the paper.  

\subsection{Supersymmetric solutions to the Killing spinor equation}

We go beyond the probe approximation and address the question: are there any solutions for the metric and 4-form field strength within the above ansatze which satisfy the Killing spinor equation for 11-dimensional supergravity:
\be
\delta_{\epsilon}\Psi_{A} = (\partial_{A}  + \frac{1}{4} \omega_A^{ab} 
\hat{\Gamma}_{ab}
+ \frac{1}{144}{\Gamma_{A}}^{BCDE}F_{BCDE}
-\frac{1}{18}\Gamma^{BCD}F_{ABCD})\epsilon = 0
\label{susy}
\ee
where $\epsilon$ is given by (\ref{epsilon}).  
The requirement $\delta_{\epsilon} \Psi_A = 0$ can be expressed as independent equations which give constraints on $F$, $g, B, \Omega$. 

\no
The constraints determine $F$ as well as impose conditions on $H_1, H_2$, $B$ and $\Omega$.  The 4-form field strength is found to be:
\bea
F &=& -\frac{3}{16}H_2\partial_\alpha \ln H_1\epsilon_{\alpha\beta\delta}(\Omega + \bar{\Omega})\wedge dx^\beta\wedge dx^\delta \nonumber \\
&+& \frac{1}{8}H_2\Omega_K^{\;\;\; J}P_{-L}^{\;\;\;\;\; I}\partial_\alpha g_{JI}\epsilon_{\alpha\beta\delta} dy^K\wedge dy^L\wedge dx^\beta\wedge dx^\delta \\
&+& \frac{3}{2}H_2^3(\Omega +\bar{\Omega})_K^{\;\;\; I}\partial_I\ln H_1 dy^K\wedge dx^8\wedge dx^9\wedge dx^{10} \nonumber \label{F1}
\eea
where $P_+$ and $P_-$ are projection operators defined as follows
\be
(P_+)_M^{\;\;\;N} = \frac{1}{2} (\delta_M^N + B_M^{\;\;\;N}) \;\;\;\;\;\;\;\;
(P_-)_M^{\;\;\;N} = \frac{1}{2} (\delta_M^N - B_M^{\;\;\;N})
\ee
Hence $P_+$ projects onto tensors of type (1,0) and $P_-$ onto those of type (0,1).  $\epsilon_{\alpha\beta\delta}$ is a completely anti-symmetric symbol with $\epsilon_{89(10)} =1$.  

\no
In addition we get a number of relations.  One set relates $H_1$ and $H_2$:
\bea
\partial_\alpha\ln (H_2H_1^2) &=&0 \\ \nonumber
\partial_I\ln (H_2H_1^2) &=&0. 
\eea
These equations determine, up to a constant which can be absorbed in a coordinate redefinition, that:
\be
H_1^{-6} = H_2^{3} \equiv H.
\ee
There are also a set of relations involving derivatives along the overall transverse directions:
\bea
0 &=& B^{IJ}\partial_\alpha\Omega_{IJ} =  B^{IJ}D_\alpha\Omega_{IJ} \nonumber \\
0 &=& \eta_{m\bar{n}}(e^{mI}\partial_\alpha e^{\bar{n}}_I - e^{\bar{n}I}\partial_\alpha e^m_I ) = (\Omega^{IJ}\partial_\alpha\bar{\Omega}_{IJ} - \bar{\Omega}^{IJ}\partial_\alpha\Omega_{IJ}) \nonumber \\
0&=& \partial_\alpha g_{MK} - B_K^{\;\;\;I}B_M^{\;\;\; J}\partial_\alpha g_{IJ} + 2g_{MK}\partial_\alpha\ln H_1 \label{alpha}
\eea
The last of these three equations implies a relation between $H$ and the determinant of the metric $g$:
\be
0 = \partial_\alpha\ln (H^{-2/3}\det g).
\ee 
In addition, there are these further conditions which involve derivatives along the Calabi-Yau:
\bea
\label{I} 
0 &=& \frac{1}{2}(\Omega_{JK}\partial_IB^{JK} - 2\Omega^{JK}\partial_J g_{KI}) + \Omega_I^{\;\;\; J}\partial_J\ln H_1 + 3\bar{\Omega}_I^{\;\;\; J}\partial_J \ln H_1 \nonumber \\
&=& \frac{1}{2}B^{JK}D_I\Omega_{JK} + \Omega_I^{\;\;\; J}\partial_J\ln H_1 + 3\bar{\Omega}_I^{\;\;\; J}\partial_J \ln H_1 \nonumber \\
0 &=& \Omega^{JK}\partial_I\bar{\Omega}_{JK} - \bar{\Omega}^{JK}\partial_I{\Omega}_{JK} - 8B^{JK}\partial_Jg_{KI} - 16B_I^{\;\;\; J}\partial_J\ln H_1 \\
&=& \Omega^{JK}D_I\bar{\Omega}_{JK} - \bar{\Omega}^{JK}D_I{\Omega}_{JK}  - 16B_I^{\;\;\; J}\partial_J\ln H_1  \nonumber
\eea

\no
Using (\ref{alpha}) one can derive a simple expression for $*F$, the Hodge dual of $F$:
\bea
*F &=&  \frac{1}{4}dx^0\wedge dx^1\wedge dx^2\wedge dx^3\wedge[\partial_\alpha (H^{-2/3}(\Omega + \bar{\Omega}))\wedge dx^\alpha \nonumber \\
&+& d_4(H^{-2/3}(\Omega + \bar{\Omega})) - H^{-1}d_4(H^{1/3}(\Omega + \bar{\Omega}))]
\eea  
where $d_4$ is the exterior derivative along the two-fold.  We have collected terms in a suggestive way in the above expression so that it is clear that imposing $d*F=0$ implies:
\be
d_4(H^{1/3}(\Omega + \bar{\Omega}) )= 0. \label{magic}
\ee
This equation is independent of (\ref{I}).   Once (\ref{magic}) is imposed, $*F$ can finally be expressed as:
\be
*F = d[\frac{1}{4}H^{-2/3}dx^0\wedge dx^1\wedge dx^2\wedge dx^3\wedge (\Omega + \bar{\Omega})] \label{starf}
\ee
which agrees with our expectations from considerations based on generalized calibrations \cite{kastortraschen}.

Once (\ref{magic}) is taken into account, the constraints in (\ref{I}) can be re-expressed in a manner that better illustrates their meaning, as follows:
\bea
d_4\Omega &=& -\frac{1}{12}d_4\ln H\wedge (\Omega + 3\bar{\Omega}) \nonumber \\
d_4(H^{-1/6}B) &=& 0 \label{relations}
\eea
The first of these equations shows that $d_4\Omega\in \Lambda^{(2,1)}\oplus\Lambda^{(1,2)}$.  If the almost complex structure had been integrable (i.e. if there was a complex structure compatible with the chosen almost complex structure) then $d_4\Omega$ would at most have a component in $\Lambda^{(2,1)}$.  Since in general (i.e. whenever $dy^JP_{+J}^{\;\;\;\;\; I}\partial_I\ln H$ is non-zero) the $(1,2)$ component is non-zero this shows that the almost complex structure is {\it not} integrable. 

In section 5 we will show that $iH^{-1/6}B$ is a calibration in the full geometry produced by the wrapped M5-branes.  This calibration gives a lower bound on the mass of M2-branes ending on the M5-branes.  The second equation in (\ref{relations}) occurs naturally in that interpretation.

\section{Transforming Special Lagrangians to Holomorphic cycles}
An important property of Calabi-Yau two-folds is that they are not only K{\" a}hler but hyper-K{\"a}hler.  This property means that the manifold is K{\"a}hler 
with respect to mutually incompatible complex structures, which are 
connected to each other by $SO(3)$ rotations.  This characteristic of Calabi-Yau two-folds has striking consequences and will provide us with not only an important test of our methods hitherto but will also allow us to discuss properties of previously known solutions that have not been discussed in the literature so far, at least to our knowledge. 

As emphasized we are studying branes wrapped on Special Lagrangian cycles with respect to a {\it specific} complex structure.  In general, a different complex structure will not be compatible with the Special Lagrangian character of the cycle in question.  

In our analysis so far we picked a complex structure - one with respect to which our 2-cycle is Special Lagrangian.  We then assumed that we inherit an almost complex structure from the underlying Calabi-Yau manifold; that is, the almost complex structure survives the modifications of the geometry that result from the presence of the wrapped brane.  

In this section we will explore what happens to the geometry if we pick a different complex structure - one with respect to which the cycle is holomorphic.  Quite generally the inherited almost complex structure transforms as follows under an $SU(2)$ rotation:
\bea
{e'}^u &=& \alpha e^u + \beta e^{\bar{v}} \nonumber \\
{e'}^{\bar{v}} &=& -\beta^*e^u + \alpha^*e^{\bar{v}}
\eea
where the primed ${e'}^m$ are (1,0) forms with respect to the new almost complex structure, while ${e'}^{\bar{m}}$ are (0,1) forms.   The complex functions $\alpha, \beta$ satisfy $|\alpha|^2 + |\beta|^2=1$.  As one can see these transformations mix (1,0) and (0,1) forms with respect to the original almost complex structure, thus the old and the new almost complex structures are not compatible with each other. 

Consider the above transformation with $\alpha = 1/\sqrt{2}, \beta = -i/\sqrt{2}$, 
\bea
{e'}^u &=& \frac{1}{\sqrt{2}}( e^u -i e^{\bar{v}}) \nonumber \\
{e'}^{v} &=& \frac{1}{\sqrt{2}}(ie^{\bar u} + e^{v}).
\eea
In the remainder of this section we will study the geometry of the previous section in this new almost complex structure.  We define to this end
\bea
B' &\equiv& \frac{1}{2}(e'^u\wedge e'^{\bar u} + e'^v\wedge e'^{\bar v} )\nonumber \\
\Omega ' &\equiv& e'^u\wedge e'^v
\eea
the (1,1) and (2,0) forms, respectively, with respect to this new almost complex structure.  These quantities can be expressed in terms of $B$ and $\Omega$ from the previous section:
\bea
B' &=& \frac{i}{2}(\Omega +\bar{\Omega}) \nonumber \\
\Omega ' + \bar{\Omega}' &=& 2iB \nonumber \\
\Omega ' - \bar{\Omega}' &=&\Omega - \bar{\Omega} \nonumber
\eea
Thus the role of the real part of $\Omega$ and $B$ is switched under the transformation.  We chose this particular transformation since $B$ - the would-be K{\"a}hler form - is the generalized calibration appropriate for holomorphic cycles, while the real part of $\Omega$ is the generalized calibration for Special Lagrangian cycles.  The metric defined by the frame fields are identical $g'_{IJ} = g_{IJ}$ as would have to be the case for this transformation to be sensible. 

Using the constraints (\ref{relations}) and (\ref{magic}) from the previous section we can derive a set of equations satisfied by $B$ and $\Omega$.  First, consider 
\bea
d_4\Omega' &=& d_4[\frac{1}{2}(2iB + \Omega - \bar{\Omega}) \\
&=& \frac{i}{6}B\wedge d\ln H + \frac{1}{12}d\ln H\wedge (\Omega - \bar{\Omega}) \nonumber \\
&=& \frac{1}{6}\Omega ' \wedge d\ln H, \label{omegaprime}
\eea
here, unlike the expression (\ref{relations}), the (1,2) piece is absent.  In fact, this implies that the Nijenhuis tensor is identically zero and hence the almost complex structure is integrable.  In other words, there are a set of holomorphic coordinates which can be defined on the entire manifold satisfying the usual compatibility conditions on intersecting coordinate patches.  The metric in these coordinates is given by $g'_{M\bar{N}} = B'_{M\bar{N}}$, while all other components vanish identically.  Equation (\ref{omegaprime}) can also be written as:
\be
d_4(H^{-1/6}\Omega') = 0. \label{calib}
\ee
Which, in light of the fact that there is now a complex structure, means that $H^{-1/6}\Omega'$ must be a {\it holomorphic} (2,0) form. 

Next, consider (\ref{magic}) in terms of quantities defined in the new almost complex structure:
\be
d_{4}(H^{1/3}B') = \frac{i}{2}d_4[H^{1/3}(\Omega + \bar{\Omega})] = 0.
\ee  
This equation was originally derived in \cite{danda} where it was interpreted as a K{\"a}hler condition on a re-scaled metric $G = H^{1/3}g$.  Finally $*F$ appearing in (\ref{starf}) can be re-written as:
\be
*F = d[-i\frac{1}{2}H^{-2/3}dx^0\wedge dx^1\wedge dx^2\wedge dx^3\wedge B']
\ee
which is in the form advocated in \cite{kastortraschen}.  

We have thus completely recovered the results of \cite{danda} and \cite{kastortraschen}.  In addition we have proven the correctness of the key assumption of these works: that there is a complex structure.  In \cite{danda} and \cite{kastortraschen} no use was made of the (2,0) form.  Our analysis in this paper suggests that the (2,0) form is a well-defined object in the theory which satisfies the identity (\ref{omegaprime}).    This (2,0) form will be important in defining calibrations for intersecting M2-branes, as discussed in the next section.  

\section{Calibrations in wrapped M5 background}
Calibrating forms, or calibrations, give us a method for determining the stability of brane configurations.  They provide a lower bound for volume forms and allow us to unequivocally establish the status of submanifolds as minimal volume.  Calibrating forms come in different varieties for different types of manifolds.  For instance, Calabi-Yau three-folds have calibrations associated with Special Lagrangian 3-cycles as well as even dimensional holomorphic cycles.  The calibrating forms are the unique (3,0) form for the 3-cycles, the K{\" a}hler form $\omega$ for 2-cycles and $\omega\wedge\omega$ for 4-cycles.  

In the case at hand the underlying Calabi-Yau two-fold can have two distinct families of calibrating two-forms, associated with holomorphic and Special Lagrangain 2-cycles.  Just as before, holomorphic 2-cycles are calibrated by the K{\"a}hler form $\omega$ while Special Lagrangian 2-cycles are calibrated by the (2,0) form (up to a constant phase).  

We now want to go beyond the geometry of the Calabi-Yau and consider calibrations in the full geometry of M5-branes  wrapping 2-cycles\footnote{Our methods here are closely related to those of \cite{mikhailov,  heningsonyi} who studied the problem of M5-brane/M2-brane intersections in flat space with no "backreaction"}.  To this end we consider introducing M2-branes whose spatial world volumes lie entirely inside the 4-dimensional part that was once the Calabi-Yau - i.e. the space spanned by the coordinates $y^I$. In what follows, we will denote the spatial part of the M2-brane worldvolume by $\Lambda$. 

Let $\sigma^1, \sigma^2$ be spatial world volume coordinates on the M2-brane, i.e. coordinates on $\Lambda$, and let the time coordinate coincide with the time coordinate of the background geometry produced by the wrapped M5-brane.  The induced metric on the M2-brane is then:
\be
ds^2_{M2} = H^{-1/3}dt^2+h_{ab}d\sigma^ad\sigma^b
\ee
where
\be
h_{ab} = g_{IJ}\partial_ay^I(\sigma)\partial_by^J(\sigma)
\ee
is the spatial part of the induced metric onto the worldvolume of the M2-brane.  Next we introduce the pullbacks of $\Omega$ and $B$ onto the worldvolume as:
\bea
\Omega |_{\Lambda} &\equiv& \tilde\Omega d\sigma^1\wedge d\sigma^2 \nonumber \\
B|_{\Lambda} &\equiv& \frac{1}{2}b d\sigma^1\wedge d\sigma^2
\eea
It is straightforward to prove the identity:
\be
\det h = |\tilde{\Omega}|^2 + |b|^2.
\ee
The volume of the M2-brane is then:
\be
\int\sqrt{h} d\sigma^1d\sigma^2 = \int\sqrt{|\tilde{\Omega}|^2 + |b|^2}d\sigma^1d\sigma^2
\ee

\no
It is then easy to see that a minimal volume cycle satisfies either:
\be
0=\alpha\tilde{\Omega} + \beta b \;\;\; \mbox{OR} \;\;\;  0=-\beta^*\tilde{\Omega} + \alpha^* b
\ee
where
\be
|\alpha|^2+|\beta|^2 = 1.
\ee

We know that the only allowed dynamical intersections of M5 and M2-branes are such that the intersection is one-dimensional \cite{strominger}.  In particular we know that M2-branes can end on M5-branes in such a way that the end of the M2-brane couples to a self-dual 2-form living on the M5-brane.  To make progress let us fix the almost complex structure.  For concreteness let us pick the almost complex structure so that the background M5-brane wraps a Special Lagrangian cycle.  Since $B=0$ on Special Lagrangian cycles it must be that $\tilde{\Omega}=0$, i.e. the pullback of $\Omega$ on the spatial part of the M2-brane vanishes - since the M5 and M2-brane intersection must be one-dimensional.  Thus we see that $B$ measures the volume of M2-branes in the M5-brane background.  

Notice, however, to calculate the mass of the M2-brane we must introduce a further factor of $\sqrt{g_{00}}=H^{-1/6}$ in the integral. The mass of the M2-brane is then given by the expression:
\be
m = -i\int H^{-1/6} \; b \; d\sigma^1 d\sigma^2.
\ee
where we have used $|b| = -ib$.  Since (\ref{relations}) tells us that $H^{-1/6}B$ is closed we have shown that $-iH^{-1/6}B$ is a calibration in this background.

Suppose instead that we had worked in a different almost complex structure - one with respect to which the M5-brane wrapped a holomorphic cycle.  Then the calibrating form for the M2-brane would have been $H^{-1/6}\Omega'$, giving us a natural explanation for (\ref{calib}).

\section{Summary and conclusions}
We can summarize our results as follows.  Using the methods of \cite{fh} we studied M5-branes wrapped on Special Lagrangian cycles with respect to a {\it fixed} complex structure.  We found the supergravity background corresponding to these wrapped M5-branes.  As in \cite{fh} we assumed that the directions initially corresponding to the Calabi-Yau two-fold continued to have an almost complex structure even after the geometry had been warped by the presence of the wrapped M5-brane and its associated flux. We were able to classify the full geometry by expressing the supergravity background fields in terms of a distinguised (1,1) form $B$ and a (2,0) form $\Omega$.  We showed that the almost complex structure with respect to which the cycle was Special Lagrangian was {\it not} integrable.

We then exploited the hyper-K{\"a}hler nature of the underlying Calabi-Yau by picking a new almost complex structure - one with respect to which the wrapped cycle appears holomorphic. As M5-branes wrapping holomorphic two-cycles in two-folds have been studied before, we checked that our solution using the methods of \cite{fh} yield the same results as those of \cite{danda}. However, our methods here are somewhat more powerful than those previously employed. In particular, we were able to show that if one assumes the existence of an almost complex structure, the main assumption of \cite{danda} that a complex structure exists, is in fact valid. Moreover, we showed that a closed, well-defined (2,0) form exists and we were able to construct it explicitly. We could have predicted the existence of such a form from the requirement that a calibrating form must exist corresponding to BPS states of M2-branes ending on M5-branes wrapped around holomorphic two-cycles.

Conversely, from this last insight and our ability to transform holomorphic cycles into Special Langrangians, we conclude that a closed calibrating form exists even when an M5-brane wraps a Special Lagrangian two-cycle; however, in this case the calibrating form is a (1,1) form and is given by $H^{-1/6}B$.  

Our exercise in this paper is instructive in many regards.  It gives us confidence in the methods of \cite{fh} as well as providing insights into the kinds of features of a Calabi-Yau which survive when M5-branes wrap 2-cycles as well as new structures that arise in the full geometry produced by the M5-branes. \\  

\no
{\Large \bf Acknowledgements}\\

\no
AF would like to thank Ingemar Bengtsson for discussions, the Physics Departments at Harvard University and Stockholm University, where part of this work was done, for their hospitality and Vetenskapsr{\aa}det for funds.  TZH would like to acknowledge funding from VR and to thank Cumrum Vafa and Sergei Gukov for interesting discussions regarding this and related work. IP was supported by the Wenner Gren Foundation and by the Greek Ministry of Education - E.U. program Pythagoras.

}

\begin{thebibliography}{77}
\bibitem{fh}
A.~Fayyazuddin and T.~Z.Husain {\it The Geometry of M-branes Wrapping Special
Lagrangian Cycles} hep-th/0505182
\bibitem{danda}
A.~Fayyazuddin and D.J.~Smith {\it Localized
intersections of M5-branes and four-dimensional superconformal
field theories} hep-th/9902210
\bibitem{BBS}
K.~Becker, M.~Becker and A.~Strominger,
{\it Fivebranes, Membranes and Non-Perturbative String Theory}
hep-th/9507158.
\bibitem{kastortraschen}
H.~Cho, M.~Emam, D.~Kastor and J.~Traschen
{\it Calibrations and Fayyazuddin-Smith Spacetimes}
hep-th/0009062
\bibitem{mikhailov}
A. Mikhailov, {\it BPS states and minimal surfaces}, hep-th/9708068.
\bibitem{heningsonyi}
M. Henningson and P. Yi, {\it Four-dimensional BPS spectra via M theory}, hep-th/9707251.
\bibitem{strominger}
A.~Strominger, {\it Open P-branes}, hep-th/9512059\\
P.~K.~Townsend, {\it Brane Surgery}, hep-th/9609217


\end{thebibliography}
\end{document}